\pdfoutput=1

\documentclass[11pt]{article}

\usepackage{acl}
\usepackage{amssymb}

\usepackage{times}
\usepackage{latexsym}

\usepackage[T1]{fontenc}

\usepackage[utf8]{inputenc}
\usepackage{graphicx}

\usepackage{microtype}
\usepackage{booktabs}
\usepackage{fdsymbol}

\usepackage[frozencache=true,cachedir=minted-cache]{minted} 

\title{UPV at TREC Health Misinformation Track 2021 \\
\large Ranking with SBERT and Quality Estimators}

\author{Ipek Baris Schlicht \and Angel Felipe Magnossão de Paula \and Paolo Rosso\\
        Universitat Politècnica de València, Spain}

\newif\ifproofread

\newcommand{\changemarker}[1]{%
\ifproofread
\textcolor{red}{#1}%
\else
#1%
\fi
}

\begin{document}
\maketitle
\begin{abstract}
Health misinformation on search engines is a significant problem that could negatively affect individuals or public health. To mitigate the problem, TREC organizes a health misinformation track. This paper presents our submissions to this track. We use a BM25 and a domain-specific semantic search engine for retrieving initial documents. Later, we examine a health news schema for quality assessment and apply it to re-rank documents. We merge the scores from the different components by using reciprocal rank fusion. Finally, we discuss the results and conclude with future works. 
\end{abstract}

\section{Introduction}
\proofreadfalse
People widely use Web search engines to seek health information and get medical advice on their health conditions~\cite{gualtieri2009doctor}. However, the Web hosts health misinformation. The presence of health misinformation could have adverse effects on individuals who believe in everything they read. To address this issue, this year, the TREC 2021 organized a shared task~\footnote{\url{https://trec-health-misinfo.github.io/}, accessed on 27.10.2021}. The task aims to implement an information retrieval that promotes helpful and credible information. An ideal search engine ranks credible and useful documents at the top of the ranking and ignores harmful documents. 

Health articles could present multiple aspects of medical research. Thus, the assessment of health articles is beyond fact-checking. Some medical experts and journalists develop schemes to assess the quality indicators of a health article. DISCERN~\citep{Charnock105} and the criteria from the Health News Review\footnote{\url{https://www.healthnewsreview.org/about-us/review-criteria/, accessed on 27.10.2021}} are the well-known ones. Also, health articles could contain medical terms; thus, the language models trained on the general domain would not be sufficient to encode knowledge in the texts. In this study, we exploit the criteria from the Health News Review, select the top 4 criteria that the transformers~\cite{vaswani2017attention} could perform well, and fine-tune a RoBERTa~\cite{liu2019roberta} classifier on the Health News Review dataset. We determine a reference vector that satisfies these criteria and then measure the distance between the reference vector and quality vector computed from each document using cosine similarity. Then we use the scores for ranking the documents retrieved by a BM25 model. We also implement a semantic search engine based on domain-specific sentence transformer. Finally, we merge the ranks from the different components to get fused rank lists.

\section{Task}
The input of the task is a topic which is about a health issue and medical treatment. The participants develop search engines that ideally return credible and correct information at the top of the ranking and discard incorrect information about a topic. The shared task dataset comprises two corpora: one for indexing, one for the topics. The dataset for indexing is a subset of the C4 corpus used by Google to train their T5 model\footnote{\url{https://www.tensorflow.org/datasets/catalog/c4}, accessed on 27.10.2021}. 

The NIST assessors derived query-relevance files (qrels) based on usefulness, correctness, and credibility scores~\cite{trecoverview} by using 35 topics. Briefly, these criteria are: 
\begin{itemize}
    \item Usefulness measures how much a user finds an answer useful.
    \item Correctness is computed by checking whether the answer matches the useful document.
    \item Credibility measures how credible the document is. 
\end{itemize}

\section{Methodology}
\begin{table}[]
    \centering
    \begin{tabular}{lllll}
        \toprule
         TREC ID & BM25 & SBERT & QE$^{\heartsuit}$ & QE$^{\diamondsuit}$ \\
        \midrule
          upv\_bm25 &$\blacksquare$ \\
          upv\_fuse\_2 &$\blacksquare$ & $\blacksquare$ \\
          upv\_fuse\_3 &$\blacksquare$ & & $\blacksquare$ \\
          upv\_fuse\_5 &$\blacksquare$ & & & $\blacksquare$  \\
         upv\_fuse\_7 &$\blacksquare$ & $\blacksquare$ & $\blacksquare$ \\
         upv\_fuse\_9 &$\blacksquare$ & $\blacksquare$ & & $\blacksquare$ \\
        
        \bottomrule
    \end{tabular}
    \caption{QE denotes query estimation. QE$^{\heartsuit}$ uses RoBERTa-base and QE$^{\diamondsuit}$ uses RoBERTa-large.}
    \label{tab:fusion_scores}
\end{table}
We describe our submissions\footnote{We submitted officially ten runs to TREC, we noticed that 4 of them which use a Kullback-Leibler divergence score, have an implementation issue in reranking. Therefore, we only presented the other runs in this paper.} to the track as can be seen in Table~\ref{tab:fusion_scores}. Initially, we used two search engines. The first one is BM25~\cite{DBLP:journals/ftir/RobertsonZ09} and the other one is based on a domain-specific Sentence-BERT (SBERT)~\cite{DBLP:conf/emnlp/ReimersG19}. Due to the limited resources, we indexed the search engines with the subset of the TREC corpus. This index was composed of 10k documents for each query, which were retrieved by the Pyserini implementation of BM25~\cite{DBLP:conf/sigir/YilmazCL20,DBLP:conf/sigir/LinMLYPN21}. Afterward, the ranks from the BM25 search engine were reranked with each quality estimator. In the end, in order to get the final ranks, we fused the ranks of BM25, SBERT, and the quality estimators. The following subsections describe the details of the submissions.

\subsection{Search Engines}
BM25 is a traditional retrieval method that has been widely used as a baseline retrieval system. We used the Pyserini implementation of the BM25 search engine to obtain the initial top 1000 documents for each topic. 

In addition to the BM25, we also performed a semantic search. In the semantic search, the query and the documents are embedded with an SBERT model, and then based on the cosine similarity of the embeddings, a rank list is returned for each query. Since the health articles may contain medical terms and the standard transformer models such as BERT may not be representative for these documents, we used a model pre-trained on the PubMed articles~\cite{DBLP:journals/bioinformatics/LeeYKKKSK20}. However, the model could encode only the limited number of tokens in a text. For this reason, prior to the encoding process, we first identified sentences in a document and then encoded only the first 20 sentences with the SBERT. 

\subsection{Quality Estimators for Reranking}
\begin{table*}[]
    \centering
    \begin{tabular}{lllll}
        \toprule
        \textbf{No} & \textbf{Criteria} \\
        \toprule
        1 & Does the story adequately discuss the costs of the intervention? \\
        \midrule
        2 & Does the story adequately quantify the benefits of the intervention? \\
        \midrule
        7 & Does the story compare the new approach with existing alternatives? \\
        \midrule
        8 & Does the story establish the availability of the treatment/test/product/procedure? \\
        \bottomrule
    \end{tabular}
    \caption{Criteria from Health News Review (\url{https://www.healthnewsreview.org/about-us/review-criteria/})}
    \label{tab:criteria}
\end{table*}

\begin{table*}[ht]
    \centering
    \begin{tabular}{p{0.6\linewidth}}
    \toprule
    The document is actually for advertising or marketing purposes. If so, the website might be biased or a scam designed to trick people into fake treatments or into
    buying medical products that do not live up to their claim. \\
    \midrule
    The information posted is from a personal blog or a forum, or by a non-expert person providing a medical product review or providing medical advice. Such
    subjective personal opinions or one point-of-view are considered not credible. \\
    \midrule
    The website provides or states claims that go against well-known medical consensus (e.g. smoking cigarettes does not cause cancer). \\
    \bottomrule
    \end{tabular}
    \caption{Examples of style related criteria from the TREC guidelines \url{https://trec-health-misinfo.github.io/docs/TREC-2021-Health-Misinformation-Track-Assessing-Guidelines_Version-2.pdf}}
    \label{tab:trecguid}
\end{table*}

A quality estimator is a multi-label classifier that gives a probabilistic score for each criterion. To implement a set of quality estimators, we used a publicly available dataset constructed from the Health News Review~\cite{zuo2021empirical}. According to the dataset paper, RoBERTa-based models have shown promising results on criteria 1, 2, 7, and 8 (see Table~\ref{tab:criteria}). \changemarker{For this reason, we first selected articles having  scores on these criteria. Then, we fine-tuned base and large RoBERTa models~\cite{liu2019roberta} by using the Huggingface library~\cite{wolf-etal-2020-transformers} on the dataset.} 

We hypothesize that an ideal health article would fulfill the quality criterion. Thus, we assumed a reference vector for an ideal article has 1.0 as a probability score for each criterion. Then, for each document in a topic, we estimated the quality criterion using the quality estimators and then measured the similarity score between the reference vector by calculating the cosine similarity. Finally, the documents were ranked again based on the similarity score. We repeated this process for the RoBERTa-base ( QE$^{\heartsuit}$)  and RoBERTa-large (QE$^{\diamondsuit}$) models, and we obtained two different ranks.

\subsection{Fusion of the Ranks}

Table~\ref{tab:fusion_scores} overviews the runs where $\blacksquare$ shows which method is selected to fuse. The methods leverage different metrics to retrieve or re-rank the documents. Therefore, we merge them by using reciprocal rank fusion, which ignores document scores produced the retrieval system and takes only ranks into account~\cite{cormack2009reciprocal}. We use the TrecTools~\cite{palotti2019} to run the fusion algorithm. 

\section{Evaluation}
In this section, we discuss the TREC results. The organizers provided the results from the baseline model, which is BM25. Before proceeding with evaluation, it is worth mentioning that we could not execute our models on the whole corpora due to a lack of computational resources. We obtained the results from the subset of the corpora. Therefore, we compare the results with our baseline (upv\_bm25), not the official baseline.

\begin{table}[]
    \centering
    \begin{tabular}{lll}
         \textbf{ID} & \textbf{Name} & \textbf{Metric} \\
          \textbf{1} & graded.usefulness & nDCG \\
          \textbf{2} & binary.useful-correct & nDCG\\
          \textbf{3} & binary.useful-correct* & P@10\\
          \textbf{4} & binary.useful-credible & nDCG \\
          \textbf{5} & useful-correct-credible & nDCG \\
          \textbf{6} & 2aspects.correct-credible & CAM\_MAP\\
          \textbf{7} & 2aspects.useful-credible & CAM\_MAP\\
          \textbf{8} & 3aspects  & CAM\_MAP\_3 \\
    \end{tabular}
    \caption{Mapping credibility, usefulness, and correctness evaluation to the evaluation metrics. nDCG: Normalized Discounted Cumulated
Gain, CAM: Convex Aggregating Measure}
    \label{tab:mapping}
\end{table}

\begin{table*}[]
    \centering
    \begin{tabular}{lllllllll}
        \toprule
        \textbf{Models} & \textbf{1} & \textbf{2} & \textbf{3} & \textbf{4} & \textbf{5} & \textbf{6} & \textbf{7} & \textbf{8}\\
        \midrule
        official baseline & \textbf{0.5815} & 0.3088 & \textbf{0.4279} & \textbf{0.4867} & \textbf{0.3813} & \textbf{0.1605} & \textbf{0.2357} & 0.205\\
        \midrule
        upv\_bm25 & 0.5285 & 0.3441 & 0.3828 & 0.445 & 0.3321 & 0.1452 & 0.2107 & 0.184 \\
        upv\_fuse\_2 & \textbf{0.5316} & 0.3412 & \textbf{0.3959} & 0.4413 & \textbf{0.3345} & 0.1256 & \textbf{0.2108} & \textbf{0.1858} \\
        upv\_fuse\_3 & 0.5127 & 0.2794 & 0.3666 & 0.4322 & 0.3176 & 0.12  & 0.1875 & 0.162 \\
        upv\_fuse\_5 & 0.5038 & 0.2529 & 0.3584 & 0.4296 & 0.3112 & 0.1315  & 0.1795 & 0.1539 \\
        upv\_fuse\_7 & 0.5204 & 0.2941 & \textbf{0.3835} & 0.4338 & 0.3287 & 0.1315 & 0.1964 & 0.1712\\
        upv\_fuse\_9 &  0.5185 & 0.2941 & 0.3743 & 0.4303 & 0.3173 & 0.1276 & 0.193 & 0.1675 \\
        \bottomrule
    \end{tabular}
    \caption{Credibility, usefulness, and correctness of the models regarding multiple aspects and binary relevance (The details of 1-8 are given in Table~\ref{tab:mapping}). Bold scores are better than our baseline (upv\_bm25).}
    \label{tab:results1}
\end{table*}

\begin{table}[]
    \centering
    \begin{tabular}{lll}
        \toprule
         \textbf{Models} & \textbf{Harmful} & \textbf{Helpful} \\
        \midrule
        official baseline & 0.1445 & 0.1292 \\
        \midrule
        upv\_bm25 & 0.1043 & 0.1341 \\
        upv\_fuse\_2 & 0.1084 & \textbf{0.1378} \\
        upv\_fuse\_3 & 0.1114 & 0.1093 \\
        upv\_fuse\_5 & 0.1061 & 0.1195 \\
        upv\_fuse\_7 & 0.1036 & 0.1286\\
        upv\_fuse\_9 & \textbf{0.1018} & 0.109 \\
        \bottomrule
    \end{tabular}
    \caption{Harmfulness and helpfulness of the models. Bold scores are better than our baseline (upv\_bm25).}
    \label{tab:results2}
\end{table}

The TREC evaluated the submitted runs based on the multiple aspect measures~\cite{10.1145/3121050.3121072} and the compatibility~\cite{10.1145/3340531.3411915, 10.1145/3409256.3409816}.  

Tables~\ref{tab:mapping} and \ref{tab:results1} present multiple aspect scores and binary relevance scores of the models. As expected, the models could not compete with the scores from the official baseline. When we compare the models with upv\_bm25 with the other models, upv\_fuse\_2 which combines the ranks of BM25 and SBERT has slightly performed better than the upv\_bm25. SBERT could take the words' context into account, and better estimate the similarity between sentences having domain-specific words. This improves the retrieval of useful documents. However, integrating the quality estimators has not shown a positive effect on the results, and hence, we got scores worse than our baseline. It seems likely that the four criteria might be ineffective in assessing the quality of the documents. Also, we noticed that according to TREC assessing guidelines\footnote{\url{https://trec-health-misinfo.github.io/docs/TREC-2021-Health-Misinformation-Track-Assessing-Guidelines_Version-2.pdf}, accessed on 27.10.2021}, \changemarker{quality estimation could overlap implicitly only with the criteria assessing style of documents \changemarker{as} shown in Table~\ref{tab:trecguid}. For example, a document advertising one treatment could lack of explaining the limitations of this treatment, thus, not satisfying the criteria 1,2,7 in Table~\ref{tab:criteria}. Criteria such as estimating the credibility of the web pages and the sources in the TREC guidelines are other parameters for consideration. }Therefore, future work should include these modalities into re-ranking models and support more criteria in the Health News Review guidelines. 

The other evaluation is to measure the harmfulness and helpfulness compatibility of the models. The best model should have lower harmfulness and higher helpfulness compatibility score. According to data in Table~\ref{tab:results2}, upv\_fuse\_2 is the model which has the highest helpful compatibility score. Also, integrating the QE$^{\diamondsuit}$ with the BM25 search engine and semantic search engine (upv\_fuse\_5 and upv\_fuse\_9) could slightly reduce the harmful scores. It seems that QE$^{\diamondsuit}$ could filter out a few low-quality documents, which is promising for future work. However, it also negatively influences helpfulness compatibility. As we mentioned in the previous paragraph, a potential solution is to encode more quality criteria and other aspects of credibility.

\section{Related Work}
This section overviews the related studies. 
\subsection{Quality Estimation of Health Articles}
DISCERN~\cite{Charnock105} and the Health News Review checklist are the popular schema to assess the quality of health articles. However, DISCERN is used only for articles related to disease treatments. Unlike DISCERN, the checklist of the Health News Review applies to diverse topics in the medical domain, and the related datasets are publicly available for implementing a quality estimation classifier. 

Researchers used the checklist to automize the quality estimation. \citeauthor{afsana2020automatically} (2020) are the first who investigated traditional machine learning algorithms, namely, SVM, Naive Bayes, Random Forest and Ensemble Vote classifier on Health News Review dataset. Furthermore, they designed various features such as Tf-idf, LIWC, POS tags, citations on the text. \cite{zuo2021empirical,al2020automatic} further examined the transformer models on the dataset and compared them with traditional models. Among the transformer models, RoBERTa has shown good results~\citep{zuo2021empirical}. In our paper, we integrate the RoBERTa-based classifiers as a part of the re-ranking task. 

\subsection{TREC 2020: Health Misinformation Track}
The previous Health Misinformation Track focused on retrieval of COVID-19 related documents from the Common Crawl corpus~\cite{DBLP:conf/trec/ClarkeRSMZ20}. We briefly describe the best-performing models. One approach leveraged T5 models~\cite{DBLP:journals/jmlr/RaffelSRLNMZLL20} to re-rank the documents according to their stance and relevance scores~\cite{DBLP:conf/trec/PradeepMZCXNL20}. The other model \cite{DBLP:conf/trec/LimaWAM20} estimates the credibility and misinformation score of the documents by employing classifiers and using the scores for re-ranking the documents. Although this study inspires our method to penalize misleading documents, we estimate quality using the schema designated explicitly for health documents.

\section{Conclusion}

In this paper, we have described our submissions to the TREC Health Misinformation Track 2021. First, we implemented a BM25 and a semantic search engine based on a domain-specific SBERT. Also, we used quality estimators based on RoBERTa models trained on the Health News Review dataset as a reranking model. Finally, we merged the ranks from the different components to get the final lists. The current study results show that integrating the rank lists from BM25 and the semantic search engine could improve the scores; however, the quality estimators were inefficient. As future work, we plan to study the integration of the different modalities for determining credibility and encoding the other aspects of the Health News Review guideline.

\section*{Acknowledgements}
The work of Paolo Rosso was in the framework of the MISMIS-FAKEnHATE on MISinformation \& MIScommunication in social media: FAKE news and HATE speech(PGC2018-096212-B-C31) and XAI-DisInfodemics on eXplainable AI for disinformation and conspiracy detection during infodemics (PLEC2021-007681) research projects funded by the Spanish Minitry of Science and Innovation as well as IBERIFIER, the Iberian Digital Media Research and Fact-Checking Hub funded by the European Digital Media Observatory (2020-EU-IA-0252). Also, we would like to thank Ronak Pradeep for his help in providing the subset of the index corpus to us. 
\bibliography{anthology,custom}
\bibliographystyle{acl_natbib}

\end{document}